\documentclass[12pt]{article}
\usepackage[margin=1in]{geometry}
\usepackage{graphicx}
\usepackage{subcaption}
\graphicspath{{./plots/}}
\usepackage{float}
\usepackage{amsmath}
\usepackage[none]{hyphenat}
\usepackage[hidelinks,backref=page]{hyperref}
\renewcommand*{\backref}[1]{}
\renewcommand*{\backrefalt}[4]{%
	\ifcase #1 (Not cited.)%
	\or        (Cited on page~#2.)%
	\else      (Cited on pages~#2.)%
	\fi}
\usepackage[numbers]{natbib}
\usepackage{amssymb}
\usepackage{url}
\usepackage{amsthm}
\usepackage[nottoc,notlot,notlof]{tocbibind}
\usepackage[title,titletoc]{appendix}
\usepackage{makecell}
\usepackage{enumitem}

\newtheorem{Def}{Definition}

\newtheorem*{Thm*}{Theorem}

\newcommand{\R}{\mathbb{R}}
\newcommand{\E}{\mathbb{E}}
\renewcommand{\P}{\mathbb{P}}

\DeclareMathOperator*{\argmin}{arg\,min}

\makeatletter
\newcommand*{\rom}[1]{\expandafter\@slowromancap\romannumeral #1@}
\makeatother

\makeatletter
\def\namedlabel#1#2{\begingroup
	\def\@currentlabel{#2}
	\label{#1}\endgroup
}
\makeatother

\begin{document}
	
	\title{UK Income Inequality and Taxation, 2000--2023: A $\kappa$-generalised Distribution Analysis}
	\author{Samuel Forbes}
	\date{}
	\maketitle
	
	\begin{abstract}
		We analyse the UK income distribution from 2000 to 2023 using HMRC annual percentile data for both pre-tax and post-tax income. We fit a prefactor-adjusted $\kappa$-generalised specification to the data by weighted non-linear least squares and use inverse transform sampling to generate simulated income populations. The results suggest a redistribution of income shares over the period: the bottom 40\% appears to have increased its share, the middle-upper part of the distribution (50th--90th percentiles) lost share, the top 10\% remained broadly stable, and the top 1\% increased its share of pre-tax income. Because the modified specification is defined only above a positive threshold, conclusions concerning the lower tail should be interpreted with some caution. Using simulated 2023 pre-tax incomes to examine tax reform scenarios, we find that revenue-equivalent tax increases on high-income earners must be more than four times as large as comparable increases on lower-income earners. This suggests that, despite increased concentration at the top, the UK tax base remains driven primarily by the large number of taxpayers outside the very top of the distribution.
	\end{abstract}
	
	\section{Introduction}
	
	Numerous empirical studies suggest that the distributions of income and wealth are well approximated by an exponential form over most of the population and by a power law in the upper tail \cite{clementi2023kaniadakis,clementi2016distribution,druagulescu2001exponential,forbes2022data}. The $\kappa$-generalised distribution \cite{clementi2023kaniadakis,clementi2016distribution}, which exhibits these asymptotic properties, is therefore a natural candidate for modelling income and wealth distributions.
	
	A major challenge in empirical work on income distributions, as with wealth distributions, is the lack of access to complete, reliable, and consistently defined microdata. Ideally, one would estimate a parametric model directly from individual-level observations using maximum likelihood or a related method. In practice, however, such data are often not publicly available, and empirical analysis must instead rely on more limited sources, such as grouped or percentile-based data.
	
	In this study, we use publicly available UK pre-tax and post-tax income percentile data from His Majesty’s Revenue and Customs (HMRC), covering the years 2000 to 2023, with the exception of 2009, for which data are unavailable. These percentile series are constructed by HMRC using the Survey of Personal Incomes (SPI). A detailed description of the SPI and its methodology is provided by Advani et al.~\cite{advani2023measuring}.
	
	When fitting the HMRC percentile data, we found that the standard $\kappa$-generalised specification did not provide a satisfactory fit, even under unweighted estimation. We therefore introduce a modified specification in which the tail function is multiplied by a constant prefactor exceeding one. This modification implies that the fitted specification is defined only above a positive threshold \(x_m>0\), rather than over the full support \((0,\infty)\). Accordingly, it should be interpreted as an empirical approximation to the observed percentile distribution above a lower threshold, rather than as a complete model of the entire income distribution. The reason this adjustment is required is not yet clear. A plausible explanation is that the discrepancy reflects the use of grouped percentile data rather than individual-level observations, although differences in fitting methodology may also play a role. We leave a fuller investigation of this issue to future work.
	
	We fit the prefactor-adjusted $\kappa$-generalised specification to the percentile data using weighted non-linear least squares. The fitted distributions are then used to generate simulated income populations for each year, from which we infer changes in inequality over time. We compute the Gini coefficient and the Theil index and find that both measures remain broadly stable over the sample period, apart from a noticeable increase during the global financial crisis. This relative stability is consistent with previous studies \cite{atkinson2020different,jenkins2022getting}, although the thresholded nature of the fitted specification means that inequality measures sensitive to the lower tail should be interpreted with some caution.
	
	To obtain a more detailed picture of distributional change, we calculate income shares, defined as the proportion of total income accruing to specified quantile groups. The fitted distributions suggest that, particularly since the financial crisis, lower-income groups have experienced a modest increase in income share, while upper-middle groups have experienced a modest decline. Because the modified specification is defined only for incomes exceeding \(x_m>0\), these lower-tail results should be interpreted cautiously. By contrast, conclusions concerning upper-income shares are likely to be more robust.
	
	We therefore examine top income shares by calculating the shares of the top 5\%, 1\%, 0.1\%, and 0.01\% for both pre-tax and post-tax income. The pre-tax share of the top 5\% remains broadly stable, while the shares of the top 1\% and top 0.1\% increase slightly over the sample period. Taken together, these findings suggest a mild compression across much of the distribution, with income shares outside the very top tending to converge. Relative to the estimates reported by Advani et al.~\cite{advani2023measuring}, our estimated top shares are somewhat lower, by several percentage points, although the qualitative trends are similar.
	
	Using the 2023 simulated income distribution, we then examine how changes in tax rates within a simplified bracketed tax system affect the tax share, defined as the ratio of total tax revenue to total income. Whereas HMRC typically estimates the revenue effects of relatively small tax reforms \cite{hmrc_direct_effects_2025}, our analysis considers larger adjustments and asks which changes in tax rates yield equivalent tax shares.
	
	We find that substantial increases in the tax rate applied to the highest-income bracket are required to match the effect of comparatively modest increases in the rate applied to the broadest bracket. In particular, an increase in the top-band rate of more than four times the increase in the broadest-band rate is needed to generate an equivalent tax share, holding other bracket rates constant. For example, a 5\% increase in the general tax rate corresponds to more than a 20\% increase in the tax rate applied to the highest-income group. This suggests that, even in the presence of increasing concentration at the top, the aggregate tax base remains strongly influenced by the large number of taxpayers outside the very top of the distribution. Diamond and Saez \cite{DiamondSaez2011} provide a theoretical framework under which high top tax rates may be optimal; however, our analysis is purely empirical and does not attempt to assess tax optimality. The code used for all analyses and figures is available in the author’s GitHub repository \cite{saf_GitHub}.
	
	\section{Fitting the Income Distribution}
	
	\subsection{Data}
	
	Let \(X \in (0,\infty)\) denote income. The data we analyse are income percentiles
	$$
	\{x_1,x_2,\dots,x_{99} : x_1<x_2<\dots<x_{99}\},
	$$
	such that
	$$
	\P(X \le x_i)=\frac{i}{100}
	\quad \Leftrightarrow \quad
	\P(X>x_i)=1-\frac{i}{100},
	\qquad i=1,2,\dots,99.
	$$
	
\begin{figure}[H]
		\centering
		\captionsetup{justification=centering}
		\includegraphics[width=\textwidth]{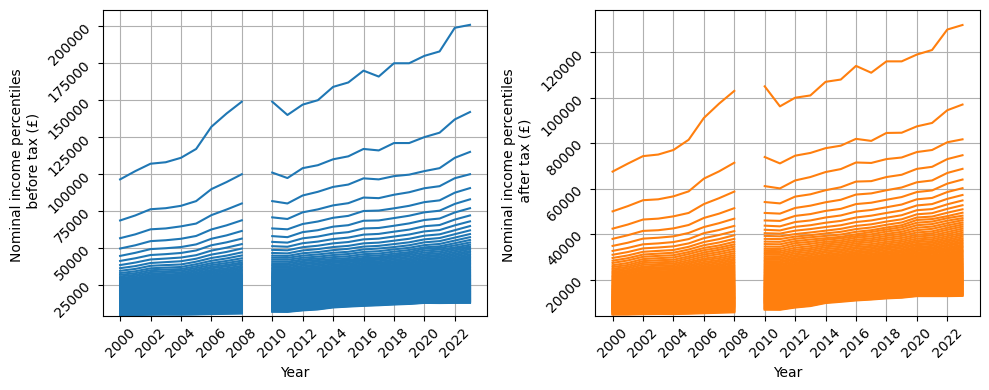}
		\caption{Nominal UK pre-tax and post-tax income percentiles, 2000--2023 (excluding 2009). Source: HMRC \cite{govuk_percentile_income_before_after_tax_2025}.}
		\label{nominal_income_percentiles}
	\end{figure}

	The HMRC UK pre-tax and post-tax income percentiles for the years 2000--2023, excluding 2009, are plotted in Figure~\ref{nominal_income_percentiles}.
	
	\subsection{\boldmath{$\kappa$}-generalised Distribution}
	
	We now define the $\kappa$-generalised distribution \cite{clementi2016distribution,clementi2023kaniadakis}, originating in statistical physics \cite{kaniadakis2001_non_linear_kinetics}, together with the modified specification used in this paper. Many properties of the distribution, including its moments and associated inequality measures, are discussed in \cite{clementi2016distribution,clementi2023kaniadakis}. Here we summarise the key definitions and tail properties.
	
	\begin{Def}[\textbf{$\kappa$-exponential and $\kappa$-logarithm} \cite{clementi2016distribution,clementi2023kaniadakis}]
		The $\kappa$-exponential is defined by
		\begin{equation}
			\exp_{\kappa}(x)=\left(\sqrt{1+\kappa^2x^2}+\kappa x\right)^{1/\kappa},
			\qquad x\in\R,\ \kappa>0.
			\label{k_exp}
		\end{equation}
		Its inverse, the $\kappa$-logarithm, is defined by
		\begin{equation}
			\log_{\kappa}(x)=\frac{x^{\kappa}-x^{-\kappa}}{2\kappa},
			\qquad x>0,\ \kappa>0.
		\end{equation}
	\end{Def}
	
	The $\kappa$-exponential has the following relevant asymptotic properties\footnote{Here \(f(x)\underset{x\to a}{\sim} g(x)\) means \(\lim_{x\to a} f(x)/g(x)=1\).} \cite{clementi2016distribution,clementi2023kaniadakis}:
	\begin{equation}
		\exp_{\kappa}(x)\underset{x\to 0}{\sim}\exp(x),
		\qquad
		\exp_{\kappa}(x)\underset{x\to -\infty}{\sim}( -2\kappa x)^{-1/\kappa}.
		\label{k_gen_ass}
	\end{equation}
	
	\begin{Def}[\textbf{$\kappa$-generalised income distribution} \cite{clementi2016distribution,clementi2023kaniadakis}]
		For \(\alpha,\beta>0\) and \(\kappa\in(0,1)\), the $\kappa$-generalised income distribution is defined by the survival function
		\begin{equation}
			\P(X>x)=\exp_{\kappa}(-\beta x^{\alpha}),
			\qquad x>0.
			\label{k_gen_dist_tail}
		\end{equation}
	\end{Def}
	
	By \eqref{k_gen_ass}, it follows that
	\begin{equation}
		\P(X>x)\underset{x\to 0^+}{\sim}\exp(-\beta x^{\alpha}),
		\qquad
		\P(X>x)\underset{x\to \infty}{\sim}(2\kappa\beta)^{-1/\kappa}x^{-\alpha/\kappa}.
		\label{k_gen_assymptotes}
	\end{equation}
	Thus, the $\kappa$-generalised income distribution is asymptotically Weibull at low income levels and asymptotically power law in the upper tail.
	
	In fitting the HMRC percentile data, we found that a satisfactory fit required modifying the $\kappa$-generalised tail function by introducing a prefactor greater than one.
	
	\begin{Def}[\textbf{Modified $\kappa$-generalised specification}]
		For \(\Delta>1\), \(\alpha,\beta>0\), and \(\kappa\in(0,1)\), define
		$$
		x_m=\left(-\frac{1}{\beta}\log_{\kappa}\!\left(\frac{1}{\Delta}\right)\right)^{1/\alpha}.
		$$
		For \(x>x_m\), the modified $\kappa$-generalised specification is given by
		\begin{equation}
			\P(X>x)=\Delta \exp_{\kappa}(-\beta x^{\alpha}).
			\label{k_gen_dist_mod_tail}
		\end{equation}
	\end{Def}
	
	The standard $\kappa$-generalised distribution is defined over all incomes greater than \(0\), whereas the modified specification is defined only for incomes exceeding \(x_m>0\). Thus, provided that \(x_m\) is relatively small, the modified specification fits most of the observed income distribution, though not its entirety. The moments and asymptotic behaviour of the modified specification differ from those of the standard version only by the multiplicative factor \(\Delta\).
	
	\subsection{Fitting}
	
	To place greater emphasis on the upper tail of the income distribution, we use weighted non-linear least squares, with larger weights assigned to higher income percentiles. Specifically, we choose weights proportional to inverse tail probability raised to a power \(\gamma \ge 0\). Let \(\P_{\Theta}(X>x_i)\) denote the modified $\kappa$-generalised tail in \eqref{k_gen_dist_mod_tail}, with parameter vector
	$$
	\Theta=\{\Delta,\kappa,\alpha,\beta\}.
	$$
	The weighted non-linear least squares model is
	\begin{equation}
		\P(X>x_i)=\P_{\Theta}(X>x_i)+\varepsilon_i,
		\qquad
		\E[\varepsilon_i]=0,
		\qquad
		\mathrm{Var}(\varepsilon_i)=1/w_i,
		\label{wnll}
	\end{equation}
	where \(w_i\) is the weight assigned to percentile \(x_i\), for \(i=1,2,\dots,99\).
	
	We estimate \(\Theta\) numerically by minimising the weighted sum of squares using the \texttt{curve\_fit} function from SciPy \cite{scipy_curve_fit}. Writing the fitted parameter vector as
	$$
	\hat{\Theta}=\{\hat{\Delta},\hat{\kappa},\hat{\alpha},\hat{\beta}\},
	$$
	we solve
	\begin{align*}
		\hat{\Theta}
		&=\argmin_{\Theta}\sum_{i=1}^{99} w_i\big(\P(X>x_i)-\P_{\Theta}(X>x_i)\big)^2 \\
		&=\argmin_{\Theta}\sum_{i=1}^{99} w_i\left(\left(1-\frac{i}{100}\right)-\P_{\Theta}(X>x_i)\right)^2.
	\end{align*}
	
	We choose the weights as
	$$
	w_i=\frac{1}{\P(X>x_i)^{\gamma}}
	=\frac{1}{(1-i/100)^{\gamma}}.
	$$
	
	If \(\gamma=0\), then \(w_i=1\) for all \(i\), and the procedure reduces to ordinary non-linear least squares. In practice, this gives a better fit to the lower percentiles but a poorer fit to the upper-tail behaviour. For \(\gamma>0\), the weights increase with percentile rank, improving the fit to the highest incomes at the cost of a weaker fit at the lower end. We set \(\gamma=1.3\) on the basis of experimentation across different values of \(\gamma\). A more systematic choice of weights would be a useful direction for future work.
	
\begin{figure}[H]
		\centering
		\captionsetup{justification=centering}
		\includegraphics[width=0.5\textwidth]{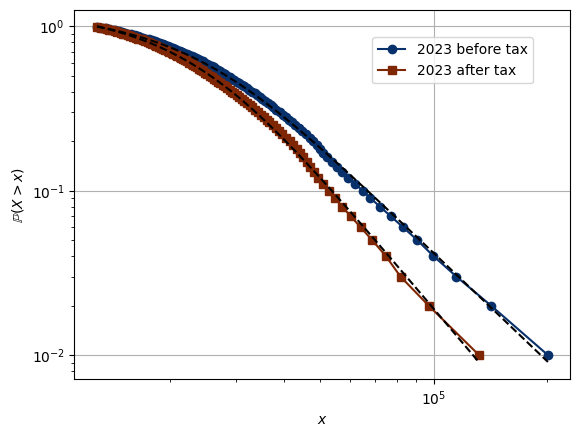}
		\caption{Income fits (black dashed lines) to UK pre-tax and post-tax income percentiles for 2023 using the modified $\kappa$-generalised specification.}
		\label{income_dist_2023}
	\end{figure}

\begin{figure}[H]
		\centering
		\captionsetup{justification=centering}
		\includegraphics[width=0.7\textwidth]{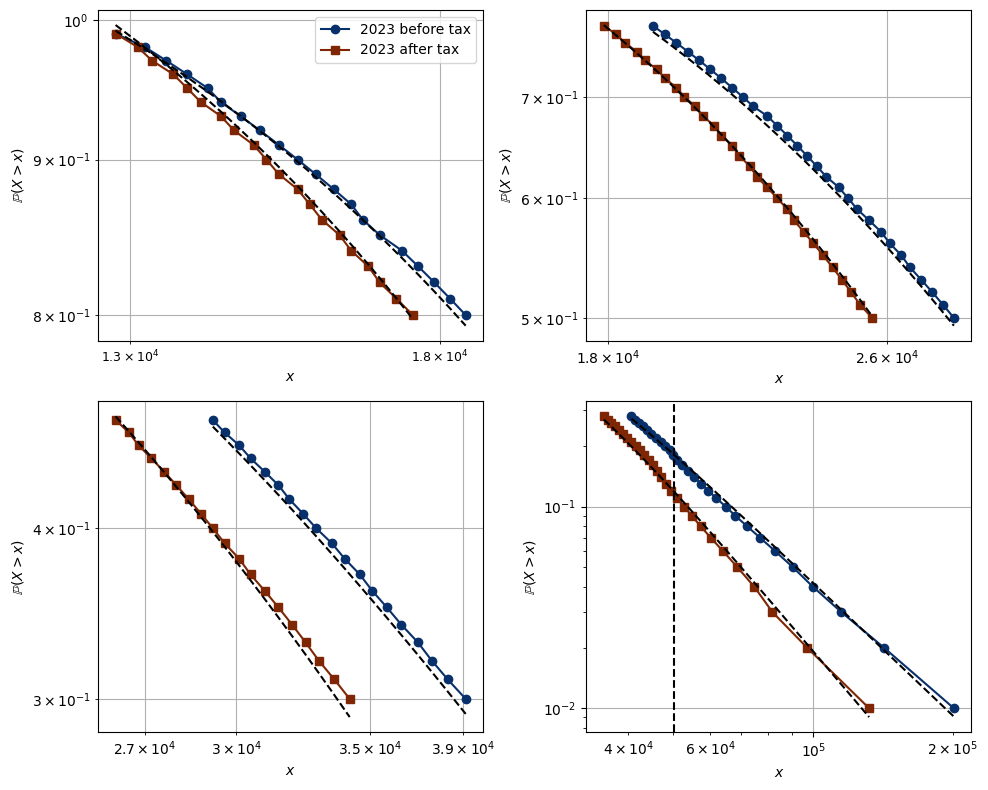}
		\caption{Income fits (black dashed lines) to UK pre-tax and post-tax income percentiles for 2023 using the modified $\kappa$-generalised specification, shown over four income intervals.}
		\label{income_dist_break_down_2023}
	\end{figure}

	Figure~\ref{income_dist_2023} shows the fitted pre-tax and post-tax income percentiles for 2023. Figure~\ref{income_dist_break_down_2023} presents the same fits over four income intervals. The fit is not exact, and the residual pattern is clearly non-random: the first few percentiles are overpredicted, middle to upper-middle incomes are underpredicted, and the highest incomes (apart from the 99th percentile) are overpredicted. Figure~\ref{income_dist_fits} in the appendix shows the fitted curves for all years from 2000 to 2023, excluding 2009. Tables~\ref{tab:annual_parameters_bt} and \ref{tab:annual_parameters_at} report the estimated parameter values.
	
	\section{Inverse Sampling}
	
	Using the fitted parameters \(\hat{\Theta}\), we generate simulated income values by inverse transform sampling from the fitted modified specification. For \(p \in (0,1)\) drawn uniformly at random, inversion of \eqref{k_gen_dist_mod_tail} gives
	\begin{equation}
		s = \left(-\frac{1}{\hat{\beta}} \log_{\hat{\kappa}}(p/\hat{\Delta})\right)^{1/\hat{\alpha}},
		\label{k_gen_mod_sample}
	\end{equation}
	which yields a simulated income value satisfying \(s>x_m\).
	
	Repeating this procedure independently produces a simulated income population
	\begin{equation}
		\{s_1, s_2, \dots, s_n\},
		\label{sample1}
	\end{equation}
	where \(n = 10^6\) is taken to be sufficiently large to yield stable estimates. These simulated populations are then used to estimate inequality measures and to study the implications of alternative tax schedules.
	
	
	\section{Inequality}
	
	To analyse inequality, we compute inequality measures using the simulated population for each year. We first sort the sample in \eqref{sample1} so that
	$$
	s_1 \le s_2 \le \dots \le s_n.
	$$
	
	The inequality measures \cite{advani2023measuring,cowell2011measuring} used in this paper are as follows.
	
	\begin{enumerate}
		\item \textbf{Gini index} \(G\), which measures the average difference between pairs of incomes. It ranges from \(0\) under perfect equality to \(1\) under maximal inequality. We use the formula
		\begin{equation}
			G=\frac{2\sum_{i=1}^{n} i s_i}{n\sum_{i=1}^{n} s_i}-\frac{n+1}{n}.
		\end{equation}
		
		\item \textbf{Theil index} \(T\), an entropy-based inequality measure originating in information theory. It takes the value \(0\) under perfect equality and increases with inequality:
		\begin{equation}
			T=\frac{1}{n}\sum_{i=1}^{n}\frac{s_i}{\mu}\log\left(\frac{s_i}{\mu}\right),
			\qquad
			\mu=\frac{1}{n}\sum_{i=1}^{n} s_i.
		\end{equation}
		
		\item \textbf{Income share} \(S_A\), defined as the share of total income accruing to a group \(A\subseteq\R\):
		\begin{equation}
			S_A=\frac{\sum_{s_i\in A} s_i}{\sum_{i=1}^{n} s_i},
			\qquad A\subseteq\R.
			\label{income_share}
		\end{equation}
		
		\item \textbf{Power-law coefficient} \(C\), the asymptotic tail exponent in \eqref{k_gen_assymptotes}:
		\begin{equation}
			C=\alpha/\kappa.
		\end{equation}
		Smaller values of \(C\) correspond to heavier upper tails and therefore greater concentration at high incomes.
	\end{enumerate}
	
\begin{figure}[H]
		\centering
		\captionsetup{justification=centering}
		\includegraphics[width=\textwidth]{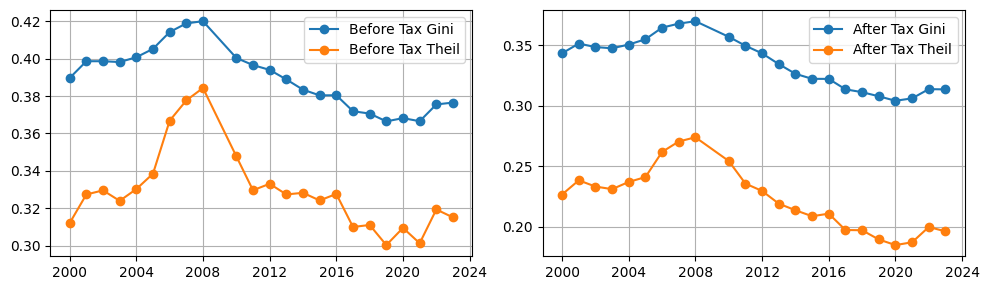}
		\caption{Gini and Theil indices of the simulated pre-tax and post-tax income populations over the years 2000--2023, excluding 2009.}
		\label{Gini_Theil}
	\end{figure}

	Figure~\ref{Gini_Theil} shows the Gini and Theil indices over time for the simulated populations. The Gini coefficient remains broadly stable for both pre-tax and post-tax income, although there is a slight upward trend from 2000 to 2008, followed by a downward movement around the financial crisis, then relative stability thereafter. The Theil index follows a similar pattern, with a more pronounced spike around 2008 for pre-tax income. This qualitative behaviour is consistent with earlier work on UK income inequality \cite{advani2023measuring}.
	
\begin{figure}[H]
		\centering
		\captionsetup{justification=centering}
		\includegraphics[width=\textwidth]{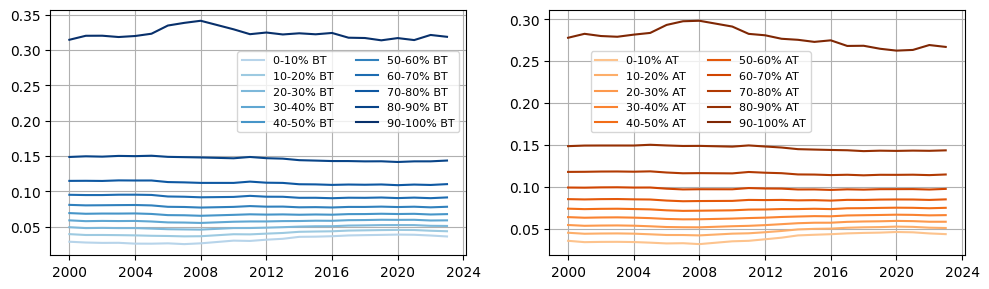}
		\caption{Decile UK pre-tax and post-tax income shares from 2000--2023, excluding 2009.}
		\label{income_decile_shares}
	\end{figure}

\begin{figure}[H]
		\centering
		\captionsetup{justification=centering}
		\includegraphics[width=\textwidth]{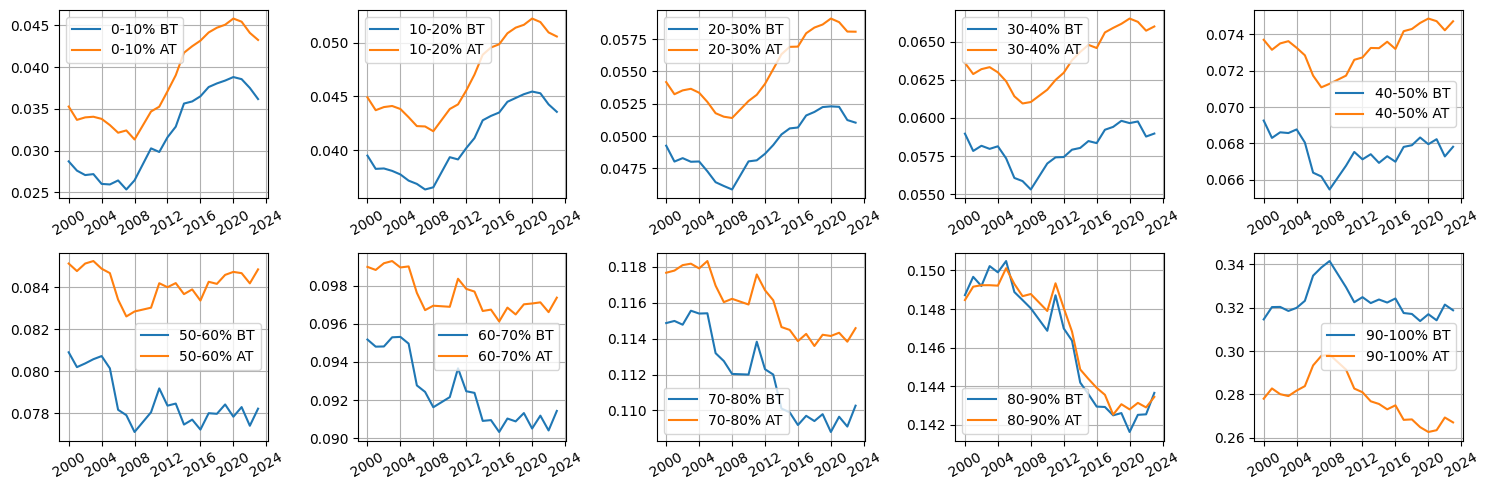}
		\caption{Decile UK pre-tax and post-tax income shares from 2000--2023, excluding 2009, shown on separate plots.}
		\label{income_decile_shares1}
	\end{figure}

	Figure~\ref{income_decile_shares} shows decile pre-tax and post-tax income shares over 2000--2023, excluding 2009. The trends are broadly stable, with some indication that lower decile shares increased modestly relative to upper decile shares. These patterns are more clearly visible in Figure~\ref{income_decile_shares1}, which presents the same series on separate plots.
	
\begin{figure}[H]
		\centering
		\captionsetup{justification=centering}
		\includegraphics[width=\textwidth]{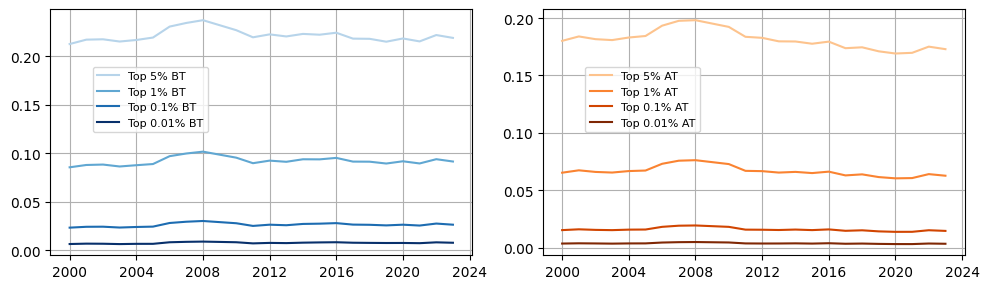}
		\caption{Top UK pre-tax and post-tax income shares from 2000--2023, excluding 2009.}
		\label{income_top_shares}
	\end{figure}

\begin{figure}[H]
		\centering
		\captionsetup{justification=centering}
		\includegraphics[width=\textwidth]{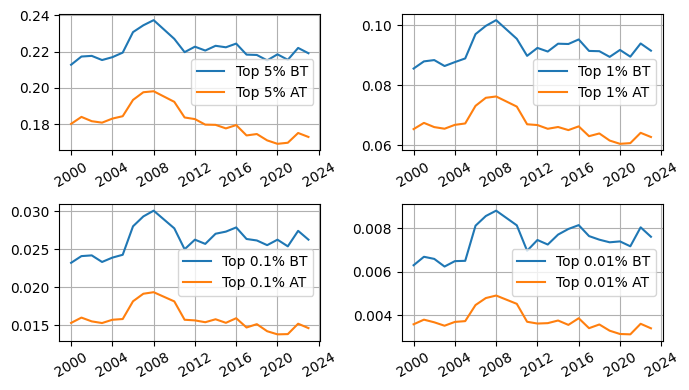}
		\caption{Top UK pre-tax and post-tax income shares from 2000--2023, excluding 2009, shown on separate plots.}
		\label{income_top_shares1}
	\end{figure}

	Figure~\ref{income_top_shares} shows the top 5\%, 1\%, 0.1\%, and 0.01\% pre-tax and post-tax income shares from 2000--2023, excluding 2009. Figure~\ref{income_top_shares1}, which separates these series into individual plots, suggests a slight increase in pre-tax income shares at the very top. By contrast, post-tax top shares are broadly stable or slightly declining. Top income shares are likely to be estimated with greater uncertainty, owing to greater volatility at the top of the distribution and to issues such as avoidance, evasion, and measurement error.
	
	As noted in the introduction, our estimated top shares are lower than those reported by Advani et al.~\cite{advani2023measuring}, but they display similar broad trends, particularly for pre-tax income.
	
\begin{figure}[H]
		\centering
		\captionsetup{justification=centering}
		\includegraphics[width=\textwidth]{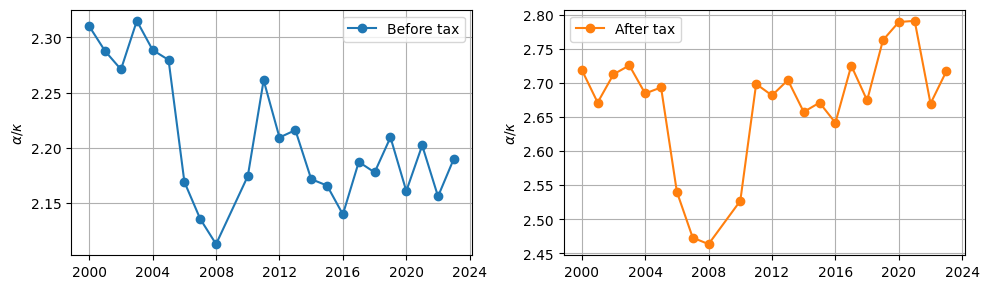}
		\caption{Estimated UK pre-tax and post-tax power-law coefficients from 2000--2023, excluding 2009.}
		\label{power_law_coef}
	\end{figure}

	Finally, Figure~\ref{power_law_coef} shows the estimated power-law coefficients for UK pre-tax and post-tax income over 2000--2023, excluding 2009. The pre-tax coefficient exhibits a gradual downward trend, consistent with increasing concentration at the top, whereas the post-tax coefficient remains comparatively stable apart from fluctuations around the financial crisis. Since this coefficient is driven by the upper tail, it is naturally most sensitive to uncertainty in top incomes.
	
	\section{Taxation}
	
	We now analyse taxation using the simulated 2023 income population. Let \(t(X)\) denote a simplified bracket tax function with four bands:
	\begin{equation}
		t(X)=
		\begin{cases}
			0, & x_m<X\le a_1, \\
			p_1(X-a_1), & a_1<X\le a_2, \\
			p_2(X-a_2)+p_1(a_2-a_1), & a_2<X\le a_3, \\
			p_3(X-a_3)+p_2(a_3-a_2)+p_1(a_2-a_1), & X>a_3.
		\end{cases}
		\label{tax_simple_fn}
	\end{equation}
	Here \(a_i>0\) and \(p_i\in[0,1]\) for \(i=1,2,3\). The first bracket \((x_m,a_1]\) is untaxed, while the brackets \((a_1,a_2]\), \((a_2,a_3]\), and \((a_3,\infty)\) are taxed at rates \(p_1\), \(p_2\), and \(p_3\), respectively.
	
	To roughly mirror the current UK system, we set the bracket cut-offs at the 1st, 85th, and 95th percentiles, which for 2023 are
	\begin{equation}
		a_1=x_1=\text{£}12{,}800,\qquad
		a_2=x_{85}=\text{£}53{,}700,\qquad
		a_3=x_{95}=\text{£}90{,}500.
		\label{a_params}
	\end{equation}
	
	We define the tax share \(R\) as the ratio of total tax revenue to total income:
	\begin{equation}
		R=\frac{\sum_{i=1}^{n} t(s_i)}{\sum_{i=1}^{n} s_i}.
		\label{tax_share}
	\end{equation}
	
	By direct calculation,
	\begin{align}
		R &= p_1 \left(S_{(a_1,a_2]} -\frac{n_1a_1-(n_2+n_3)(a_2-a_1)}{N} \right)
		+p_2 \left(S_{(a_2,a_3]} -\frac{n_2a_2-n_3(a_3-a_2)}{N} \right) \nonumber \\
		&\quad + p_3 \left(S_{(a_3,\infty]} -\frac{n_3a_3}{N} \right),
		\label{tax_share1}
	\end{align}
	where \(S_A\) is the income share of incomes in \(A\), as defined in \eqref{income_share}, and
	$$
	N=\sum_{i=1}^{n} s_i
	$$
	is total income, while
	$$
	n_1 = |\{s_i : s_i \in (a_1,a_2]\}|,\qquad
	n_2 = |\{s_i : s_i \in (a_2,a_3]\}|,\qquad
	n_3 = |\{s_i : s_i \in (a_3,\infty)\}|
	$$
	denote the numbers of simulated taxpayers in the three taxed brackets.
	
	Define
	\begin{align*}
		K_1 &= S_{(a_1,a_2]} -\frac{n_1a_1-(n_2+n_3)(a_2-a_1)}{N}, \\
		K_2 &= S_{(a_2,a_3]} -\frac{n_2a_2-n_3(a_3-a_2)}{N}, \\
		K_3 &= S_{(a_3,\infty]} -\frac{n_3a_3}{N}.
	\end{align*}
	Then \eqref{tax_share1} becomes
	$$
	R=K_1p_1+K_2p_2+K_3p_3.
	$$
	For the 2023 simulation we obtain
	\begin{equation}
		K_1=0.481,\qquad K_2=0.086,\qquad K_3=0.102,
		\label{Ks_2023}
	\end{equation}
	to three decimal places.
	
	The tax share is therefore a linear function of the bracket tax rates \(p_i\). Let \(i,j,k\) be distinct elements of \(\{1,2,3\}\). Two sets of rates \(\{p_i,p_j,p_k\}\) and \(\{\tilde{p}_i,\tilde{p}_j,\tilde{p}_k\}\) produce the same tax share when
	\begin{equation}
		K_i p_i + K_j p_j + K_k p_k
		=
		K_i \tilde{p}_i + K_j \tilde{p}_j + K_k \tilde{p}_k.
		\label{tax_shares_equal}
	\end{equation}
	
	We consider two cases.
	
	\begin{enumerate}
		\item If \(p_i=\tilde{p}_i\), then
		\begin{equation*}
			\tilde{p}_j = p_j + \frac{K_k}{K_j}(p_k-\tilde{p}_k).
		\end{equation*}
		\namedlabel{case1}{1}
		
		\item If \(\tilde{p}_k=\tilde{p}_j+y\), then
		\begin{equation*}
			\tilde{p}_j=\frac{K_i(p_i-\tilde{p}_i)+K_jp_j+K_k(p_k-y)}{K_j+K_k}.
		\end{equation*}
		\namedlabel{case2}{2}
	\end{enumerate}
	
	The proportions that make the simplified tax function \eqref{tax_simple_fn} most closely resemble the current UK tax system are
	\begin{equation}
		p_1=0.2,\qquad p_2=0.4,\qquad p_3=0.45.
		\label{p_params}
	\end{equation}
	
\begin{figure}[H]
		\centering
		\captionsetup{justification=centering}
		\includegraphics[width=\textwidth]{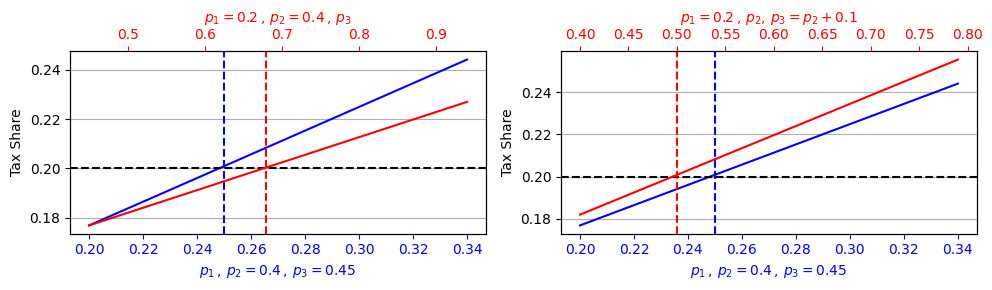}
		\caption{Tax comparisons for the 2023 simulations. In both panels, the blue curve (with blue \(x\)-axis) shows the tax share as \(p_1\) varies, with \(p_2=0.4\) and \(p_3=0.45\) fixed. \emph{Left:} the red curve (with red \(x\)-axis) shows the tax share as \(p_3\) varies, with \(p_1=0.2\) and \(p_2=0.4\) fixed. \emph{Right:} the red curve shows the tax share as \(p_2\) varies with \(p_3=p_2+0.1\) and \(p_1=0.2\) fixed. Dotted red and blue lines indicate parameter values yielding the same approximate tax share of \(0.2\), as discussed in the text.}
		\label{tax_comparisons}
	\end{figure}

	Starting from \eqref{p_params}, we first ask how much \(p_3\) must increase, holding \(p_1\) and \(p_2\) fixed, in order to generate the same tax share as an increase \(x\in[0,0.8]\) in \(p_1\), with \(p_2\) and \(p_3\) otherwise unchanged. This amounts to comparing the two sets of rates
	$$
	\{0.2+x,\,0.4,\,0.45\}
	\quad \text{and} \quad
	\{0.2,\,0.4,\,\tilde{p}_3\},
	$$
	where \(\tilde{p}_3\) is chosen so that the tax shares coincide. By Case~\ref{case1} and \eqref{Ks_2023},
	$$
	\tilde{p}_3
	= p_3 + \frac{K_1}{K_3}(p_1-\tilde{p}_1)
	= 0.45 + 4.709x.
	$$
	Thus, for fixed \(p_2=0.4\), an increase of \(x\) in \(p_1\) is equivalent, in tax-share terms, to an increase of approximately \(4.7x\) in \(p_3\). For example, increasing \(p_1\) by \(0.05\) to \(0.25\) corresponds to raising \(p_3\) to \(0.685\), with \(p_2\) held constant. This comparison is illustrated in the left panel of Figure~\ref{tax_comparisons}. The result indicates that achieving the same tax share through the highest income bracket requires substantially larger rate increases than a relatively modest change applied to the broadest taxed bracket.
	
	We now return to the baseline rates \eqref{p_params} and consider an alternative comparison. Suppose \(p_1\) is held fixed, while \(p_2\) and \(p_3\) increase together according to \(p_3=p_2+y\). We ask how much \(p_2\) must increase to generate the same tax share as before, namely that produced by an increase \(x\in[0,0.8]\) in \(p_1\) with \(p_2\) and \(p_3\) fixed. This leads us to compare
	$$
	\{0.2+x,\,0.4,\,0.45\}
	\quad \text{and} \quad
	\{0.2,\,\tilde{p}_2,\,\tilde{p}_2+y\},
	$$
	where \(\tilde{p}_2\) is chosen to equalise tax shares. Applying Case~\ref{case2} and \eqref{Ks_2023} yields
	$$
	\tilde{p}_2
	=
	\frac{K_1(p_1-\tilde{p}_1)+K_2p_2+K_3(p_3-y)}{K_2+K_3}
	=
	0.427 + 2.56x - 0.54y.
	$$
	Taking, for example, \(x=0.05\) and \(y=0.1\) gives \(\tilde{p}_2=0.5\). Hence the rates
	$$
	p_1=0.25,\qquad p_2=0.4,\qquad p_3=0.45
	$$
	produce approximately the same tax share (around \(0.2\)) as
	$$
	\tilde{p}_1=0.2,\qquad \tilde{p}_2=0.5,\qquad \tilde{p}_3=0.6.
	$$
	This comparison is shown in the right panel of Figure~\ref{tax_comparisons}.
	
	\section{Conclusion}
	
	Using UK income percentile data from 2000--2023 (excluding 2009), we fitted prefactor-adjusted $\kappa$-generalised specifications to both pre-tax and post-tax incomes, allowing us to analyse inequality trends and their fiscal implications. The fitted distributions suggest a redistribution pattern in which lower-income groups appear to have gained income share, while middle-upper earners (50th--90th percentiles) lost share, and the top 1\% increased its share of pre-tax income. Because the modified specification is defined only above a positive threshold, conclusions concerning the lower tail should be interpreted with some caution. By contrast, the evidence for relative stability in the upper distribution outside the very top, together with modest increases in the top 1\% share, is more robust. Taken together, these findings are consistent with a mild income squeeze on the middle and upper-middle parts of the distribution.
	
	Our tax analysis using the 2023 simulated income distribution yields the following approximate result: the tax-share equivalence ratio between broad-based and high-earner tax changes is greater than 1:4. Specifically, a 5\% increase in the general tax rate generates approximately the same tax share as a more than 20\% increase in the highest rate, holding other rates constant. This indicates that, despite growing concentration at the top, the UK tax base remains strongly influenced by the large number of taxpayers outside the very top of the distribution. More complex proportional changes in tax bracket rates can also generate the same overall tax share, while exhibiting a similar pattern in which higher brackets require substantially larger rate increases.
	
	These findings contribute to ongoing debates on inequality and taxation by providing empirical evidence on distributional change and on the trade-offs involved in tax reform. More broadly, the results suggest that while top income concentration matters for inequality, the revenue capacity of the tax system continues to depend heavily on the broad mass of taxpayers. A fuller assessment using individual-level sample data and maximum likelihood estimation would be a natural direction for future work.
	
	\bibliographystyle{apalike}
	\bibliography{mybib}
	
	\appendix

\section*{Appendix }

	\begin{figure}[H]
	\centering
	\captionsetup{justification=centering}
	\includegraphics[width=1.0\textwidth]{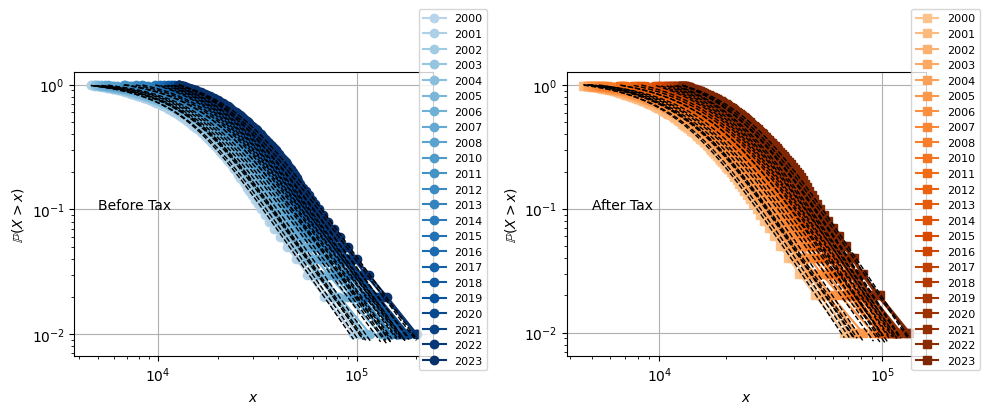}
	\caption{Income fits (black dashed lines) to before and after tax UK income percentiles using the $\kappa$-generalised distribution.}
	\label{income_dist_fits}
\end{figure}

\begin{table}[H]
	\centering
	\begin{tabular}{|c|c|c|c|c|c|c|}
		\hline
		\textbf{Year} & $\hat{x}_m$ & $\hat{ \Delta }$ & $\hat{\kappa}$ & $\hat{\alpha}$ & $\hat{\beta}$ \\
		\hline
		2000 & 4307.28 & 1.0923 & 0.8114 & 1.8743 & \(1.36 \times 10^{-8}\) \\
		2001 & 4252.75 & 1.0961 & 0.7852 & 1.7961 & \(2.79 \times 10^{-8}\) \\
		2002 & 4270.25 & 1.0766 & 0.8243 & 1.8719 & \(1.18 \times 10^{-8}\) \\
		2003 & 4424.71 & 1.0903 & 0.7732 & 1.7900 & \(2.58 \times 10^{-8}\) \\
		2004 & 4067.31 & 1.0650 & 0.8162 & 1.8680 & \(1.14 \times 10^{-8}\) \\
		2005 & 4321.59 & 1.0782 & 0.7855 & 1.7906 & \(2.33 \times 10^{-8}\) \\
		2006 & 4861.44 & 1.0931 & 0.8246 & 1.7887 & \(2.27 \times 10^{-8}\) \\
		2007 & 4653.70 & 1.0728 & 0.8634 & 1.8438 & \(1.21 \times 10^{-8}\) \\
		2008 & 5364.07 & 1.0993 & 0.8451 & 1.7855 & \(2.08 \times 10^{-8}\) \\
		2010 & 6875.29 & 1.1504 & 0.8197 & 1.7824 & \(2.03 \times 10^{-8}\) \\
		2011 & 6625.33 & 1.1512 & 0.7577 & 1.7134 & \(4.00 \times 10^{-8}\) \\
		2012 & 7611.02 & 1.1850 & 0.7826 & 1.7291 & \(3.31 \times 10^{-8}\) \\
		2013 & 8255.32 & 1.2092 & 0.7809 & 1.7305 & \(3.18 \times 10^{-8}\) \\
		2014 & 9661.04 & 1.3057 & 0.7793 & 1.6924 & \(4.84 \times 10^{-8}\) \\
		2015 & 9911.90 & 1.2774 & 0.8203 & 1.7765 & \(1.96 \times 10^{-8}\) \\
		2016 & 10561.20 & 1.3046 & 0.8292 & 1.7742 & \(1.95 \times 10^{-8}\) \\
		2017 & 11033.64 & 1.3287 & 0.8065 & 1.7640 & \(2.12 \times 10^{-8}\) \\
		2018 & 11576.99 & 1.3320 & 0.8232 & 1.7928 & \(1.50 \times 10^{-8}\) \\
		2019 & 11908.24 & 1.3375 & 0.8073 & 1.7838 & \(1.57 \times 10^{-8}\) \\
		2020 & 12638.16 & 1.3409 & 0.8529 & 1.8428 & \(8.19 \times 10^{-9}\) \\
		2021 & 12595.25 & 1.3486 & 0.8071 & 1.7773 & \(1.56 \times 10^{-8}\) \\
		2022 & 12790.21 & 1.3406 & 0.8139 & 1.7548 & \(1.84 \times 10^{-8}\) \\
		2023 & 12567.56 & 1.2698 &  0.8209 &  1.7979 & \(  1.02 \times 10^{-8} \) \\
		\hline
	\end{tabular}
	\caption{Annual income before tax parameter fits 2000 to 2023 excluding 2009}
	\label{tab:annual_parameters_bt}
\end{table}

\begin{table}[H]
	\centering
	\begin{tabular}{|c|c|c|c|c|c|c|}
		\hline
		\textbf{Year} & $\hat{x}_m$ & $\hat{ \Delta }$ & $\hat{\kappa}$ & $\hat{\alpha}$ & $\hat{\beta}$ \\
		\hline
		2000 & 4763.26 & 1.1661 & 0.6701 & 1.8221 & \(3.06 \times 10^{-8}\) \\
		2001 & 4598.62 & 1.1445 & 0.6819 & 1.8218 & \(2.87 \times 10^{-8}\) \\
		2002 & 4912.92 & 1.1475 & 0.6697 & 1.8167 & \(2.71 \times 10^{-8}\) \\
		2003 & 4989.90 & 1.1471 & 0.6676 & 1.8196 & \(2.56 \times 10^{-8}\) \\
		2004 & 5007.23 & 1.1460 & 0.6776 & 1.8190 & \(2.54 \times 10^{-8}\) \\
		2005 & 5059.16 & 1.1494 & 0.6547 & 1.7634 & \(4.10 \times 10^{-8}\) \\
		2006 & 5152.95 & 1.1343 & 0.7130 & 1.8112 & \(2.39 \times 10^{-8}\) \\
		2007 & 5446.23 & 1.1429 & 0.7337 & 1.8145 & \(2.23 \times 10^{-8}\) \\
		2008 & 5357.88 & 1.1148 & 0.7593 & 1.8706 & \(1.15 \times 10^{-8}\) \\
		2010 & 6680.02 & 1.1813 & 0.7099 & 1.7939 & \(2.30 \times 10^{-8}\) \\
		2011 & 6763.40 & 1.2089 & 0.6277 & 1.6938 & \(6.19 \times 10^{-8}\) \\
		2012 & 7629.01 & 1.2486 & 0.6316 & 1.6939 & \(5.91 \times 10^{-8}\) \\
		2013 & 8385.76 & 1.2869 & 0.6304 & 1.7047 & \(5.19 \times 10^{-8}\) \\
		2014 & 9546.98 & 1.3792 & 0.6361 & 1.6904 & \(6.06 \times 10^{-8}\) \\
		2015 & 9985.96 & 1.3930 & 0.6392 & 1.7073 & \(4.96 \times 10^{-8}\) \\
		2016 & 10598.48 & 1.4562 & 0.6275 & 1.6578 & \(8.05 \times 10^{-8}\) \\
		2017 & 11033.52 & 1.4527 & 0.6201 & 1.6901 & \(5.54 \times 10^{-8}\) \\
		2018 & 11584.61 & 1.4312 & 0.6653 & 1.7794 & \(2.12 \times 10^{-8}\) \\
		2019 & 11924.27 & 1.4779 & 0.6136 & 1.6954 & \(4.84 \times 10^{-8}\) \\
		2020 & 12683.69 & 1.5176 & 0.6007 & 1.6755 & \(5.62 \times 10^{-8}\) \\
		2021 & 12690.98 & 1.5118 & 0.5948 & 1.6601 & \(6.43 \times 10^{-8}\) \\
		2022 & 12693.45 & 1.4026 & 0.6704 & 1.7899 & \(1.54 \times 10^{-8}\) \\
		2023 & 12715.04 & 1.3391 &  0.6767 &  1.8388 & \( 8.34 \times 10^{-9} \) \\
		\hline
	\end{tabular}
	\caption{Annual income after tax parameter fits 2000 to 2023 excluding 2009}
	\label{tab:annual_parameters_at}
\end{table}

%
%

\end{document}